# Why to Decouple the Uplink and Downlink in Cellular Networks and How To Do It

Federico Boccardi[1], Jeffrey Andrews[2], Hisham Elshaer[3,4], Mischa Dohler[3]
Stefan Parkvall[5], Petar Popovski[6], Sarabjot Singh[7]

[1]Ofcom, London, UK
[2]Univ. of Texas at Austin, USA
[3]King's College London, UK
[4]Vodafone, Newbury, UK
[5]Ericsson, Stockholm, Sweden
[6]Aalborg University, Aalborg, Denmark
[7]Nokia, Berkeley, USA

23 March 2015

**Abstract**

Ever since the inception of mobile telephony, the downlink and uplink of cellular networks have been *coupled*, i.e. mobile terminals have been constrained to associate with the same base station (BS) in both the downlink and uplink directions. New trends in network densification and mobile data usage increase the drawbacks of this constraint, and suggest that it should be revisited. In this paper we identify and explain five key arguments in favor of Downlink/Uplink Decoupling (DUDe) based on a blend of theoretical, experimental, and logical arguments. We then overview the changes needed in current (LTE-A) mobile systems to enable this decoupling, and then look ahead to fifth generation (5G) cellular standards. We believe the introduced paradigm will lead to significant gains in network throughput, outage and power consumption at a much lower cost compared to other solutions providing comparable or lower gains.

1    **Introduction and Background**

From the first to the fourth generation (4G) of mobile networks, the downlink (DL) and uplink (UL) of a given communication session have been coupled: the mobile user equipment (UE) must associate with the same BS in both the DL and UL. Historically, this was a nearly optimal approach, since the strongest BS-UE connection was the same in both directions. However, this conventional approach has recently [ElsBoc14] come under scrutiny given the possible gains that can be achieved by decoupling the association in the context of a dense heterogeneous cellular network, wherein different BSs can have highly variable transmit powers and deployment topologies.

The arguments in favor of the coupled *status quo* are several. From a pure network design perspective, the logical, transport and physical channels are easier to design and operate; this pertains particularly to the synchronization of the acknowledgements (ACK/NAK), the call admission and handover procedures, DL/UL radio resource management, and power control, among others. Decoupling both links also requires strong synchronization and data connectivity (e.g. via fiber) between the BSs. From a deployment and topology perspective, until just a few years ago cellular systems have been designed and deployed under the assumption of a homogenous network with macro cells all transmitting with about the same power. From a traffic point of view, the load in both directions has been approximately the same in voice-centric 2G and early 3G systems. Moreover, 3.5G (e.g. HSPA) and 4G systems are dominated by downlink traffic, justifying the use of DL-centric association procedures rather than UL or decoupled ones.

However, a coupled association is a restrictive special case of a more general association policy without a coupling constraint. Therefore it is clear that a well-designed association policy based on Downlink/Uplink Decoupling (DUDe) can in principle outperform a coupled association. But by how much? And at what cost?

More specifically, the main questions this article attempts to answer are:

(i)     What recent trends in cellular network deployment and applications make the gains from DUDe more relevant now than in the past?
(ii)    What are the key benefits of a decoupled association in terms of throughput gain, reliability, and power conservation? How can these gains be realized in current (e.g. LTE-A) and future 5G cellular networks?
(iii)   How disruptive will these changes be to the network architecture? Are the gains large enough to be worth the trouble?

We begin with the five key arguments in favor of DUDe, and provide evidence for the corresponding gains from very recent theoretical analysis and simulation-based experiments. Then, we move on to discuss what changes will need to be made in the current and future cellular standards, and explain why, in our view, such changes are quite manageable. DUDe opens up many new interesting research questions as well, which we identify throughout the article.

## 2 Five Reasons to Decouple the Downlink and Uplink

We now articulate the five principle arguments in favor of DUDe. Our arguments are supported by a combination of recent theoretical and system-level simulation results by the present authors and others. In particular, the theoretical results are mostly sourced from the recent work [SinZhaAnd14], in which we perform a comprehensive SINR and rate analysis with DUDe in a multi-tier cellular network with spatially random UEs and BS. The UEs employ fractional UL power control and small-cell biasing is used to achieve cell range expansion: both very similar to LTE. The results are mathematical and thus transparent, albeit in some cases based on idealized models to allow tractability. We refer to this approach as the *analytical model*.

The simulation results and parameters follow largely from [ElsBoc14], and utilize an existing LTE HetNet deployment in conjunction with a high resolution 3D ray tracing channel model that takes into account clutter, terrain and building data. This ensures a highly realistic and accurate propagation model. The BS types and locations are based on Vodafone's small cell test network in the London area and consist of five macro cells covering a one kilometer square area with a dense small cell deployment embedded in the square kilometer as well. The UE distribution is based on live traffic measurements, and the UEs use the same UL fractional power control as in the analytical model. We assume that the DL association is based on the DL Reference Signal Received Power (RSRP). We refer to this approach as the *simulation model*.

As we will see, these two distinct approaches to modeling and analyzing DUDe are quite unified in terms of the conclusions they offer. Table 1 contains the cellular network notation and simulation parameters. We also use the same parameters for numerical evaluation of mathematically derived results using the analytical model.

**Table 1.** Cellular network notation and parameter values.

| *Parameter* | *UEs* | *Macro cells* | *Small cells* |
|---|---|---|---|
| Max Transmit Power | 20 dBm | 46 dBm | 30 dBm (unless otherwise specified) |
| Antenna system (simulation) | 1 Tx and 1 Rx Antenna gain = 0 dBi | 2 Tx and 2 Rx Antenna gain = 17.8 dBi | 2 Tx and 2 Rx Antenna gain = 4 dBi |
| Antenna system (analysis) | The analysis considers 1 Tx and 1 Rx isotropic antenna system for UEs, Macro cells and Small cells. | | |
| Downlink Bias | N/A | 0 dB | Varies from 0 to 8 dB |
| Spatial distribution (analytical) | Uniform | Poisson Point Process | Poisson Point Process |
| Spatial distribution (simulation) | Hotspot distribution based on realistic traffic measurements | Based on the Vodafone LTE network deployment. | Based on the Vodafone LTE small cell test network deployment. |
| Spatial Density | 330 per km$^2$ | 5 per km$^2$ | 4 small cells per macro cell |
| Channel Model | Rayleigh small scale fading | | |

| | |
|---|---|
| (analytical) | Standard path loss with exponent = 3.5 |
| | Lognormal shadowing with standard deviation = 8 dB |
| Channel Model (simulation) | 3D ray-tracing propagation model |
| Power control | Uplink fractional path loss compensation |
| Operating frequency | 2.6 GHz co-channel FDD deployment |
| Bandwidth | 20 MHz (100 frequency blocks) |
| Scheduler | Equipartition of resources among UEs (analysis) |
| | Proportional fair (simulation) |

### 2.1 Increased uplink SNR and reduced transmit power

In a typical HetNet scenario the downlink coverage area of a macro cell is much larger than that of a low power BS; indeed, this is why they are often called "small cells" BSs. The coverage area disparity is primarily attributable to the differences in downlink transmit powers, but is also due to the BS heights and antenna gains. In contrast, in the uplink all transmitters have roughly the same maximum transmit power. Therefore, a device that is associated to a macro cell in the downlink might instead wish to be associated to a small cell in the uplink, to take advantage of the reduced path loss [SmiPoGa15]. The positive effects are twofold. For UEs that are transmitting at the maximum power, a connection to a closer BS provides a higher SNR (more on the SINR below). Moreover, for a fixed target SNR, the reduced path loss alternatively allows transmit power reduction via power control.

In Figure 1, we observe the decrease in transmit power via DUDe by comparing three cases via the simulation model. The first case is the baseline with a coupled downlink/uplink association, and no small cell bias (more on bias below). The second case is still coupled, but the small cells have a 6 dB bias (which has been shown in [SinAnd14] to be a reasonable value for the bias). The third case is for DUDe. DUDe yields 2.3 dB at 50% and 3dB at 95% CDF w.r.t. the coupled association with 6 dB bias.

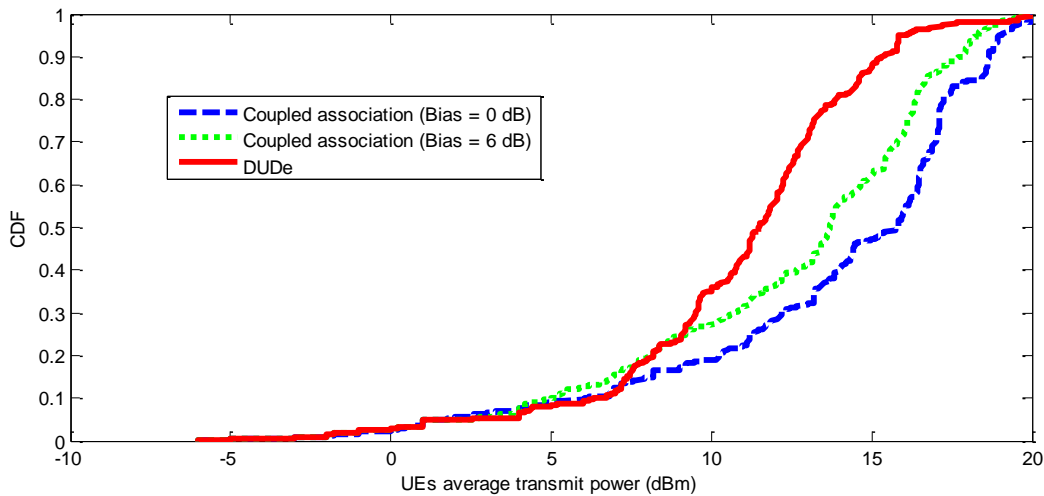

**Figure 1.** CDF of the UE's UL transmit power via the simulation model. Cell edge users (right side of figure) require higher transmit powers and thus achieve larger power reduction from DUDe.

## 2.2 Improved Uplink Interference Conditions

DUDe also decreases the uplink interference, due to multiple complementary effects.

First, and as an obvious consequence of the transmit power reduction demonstrated in the previous section, the UL interference generated to other base stations is correspondingly reduced by about 2-3 dB. This is quite significant especially for the low SINR UEs in the uplink, since at low SINR in a dense network, decreasing the interference by 3dB implies an approximate doubling of data rate.

Second, DUDe provides the ability to independently select the association that minimizes interference at both the UE and the BS. Uplink interference in a given spectral band is an aggregation of many different UEs transmissions in different cells (and possibly sectors of the same cell), as received by a given BS, say BS0. The interference generated by each of these UEs depends on its location relative to its own desired BS, the amount of power control, its distance to BS0 and the uplink precoding weights. In contrast, the downlink interference at a given UE depends on the BSs' transmit power, the downlink beamforming weights and the distance to the different BSs. On top of this, the nearly independent scheduling and loading in each of the DL and UL causes further randomness in the interference. For all these reasons, average interference levels can be quite different in the downlink and uplink resources. Therefore, a decoupled association that allows the UE (or network) to seek out the best interference environment in the two links independently can be expected to substantially outperform a coupled association, which must "split the difference".

Third, DUDe will also prove a boon for Device-2-Device (D2D) communication which, as of 3GPP Rel. 12, will take place in the uplink bands. By lowering the UL transmit power and generating less interference, DUDe will create a more benign environment for D2D receivers and thus facilitating more D2D transmissions to take place.

Finally, in addition to reducing the amount of average interference, DUDe also allows a reduction of the uplink SINR variance, as shown in Figure 2 (obtained via the simulation model) which translates into more efficient and effective UL schedulers and performance gains [AbuAliTaha14]. Specifically, w.r.t. today's best operational Case 2, the decoupling yields a reduction of 1dB on average which is about 25% at 50% CDF.

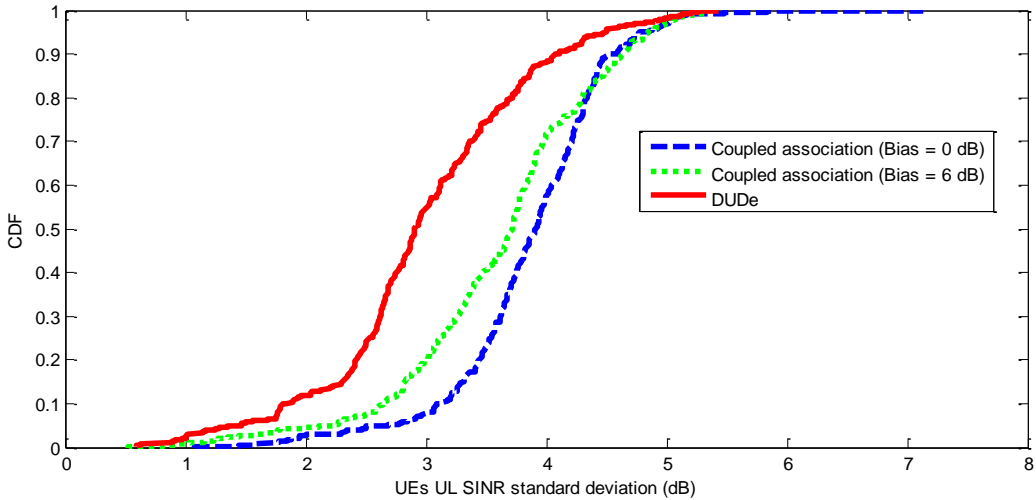

**Figure 2.** CDF of the UEs SINR standard deviation over time. DUDe reduces the variations and improves performance.

## 2.3 Improved Uplink Data Rate

Unsurprisingly, increasing the desired received power and decreasing the interference leads to higher SINR, and hence a higher spectral efficiency and data rate. However, there are additional factors which can complicate the effect of DUDe on the uplink rate.

For example, consider an LTE HetNet with small cell range expansion, which is achieved by biasing UEs towards the small cells, meaning they associate with a small cell BS even if their received power is less than from the macro cell BS (by less than the bias value). On average the optimal downlink bias is in the neighborhood of 5-10 dB as noted before, although with blanking or interference avoidance, up to 18-20 dB may be used in certain scenarios [Dam11,SinAnd14]. DL biasing leads to a better association in both directions even with coupled association, since by expanding the DL small cell coverage region, more UEs associate with the nearby small cells in the UL as well, which is also the main point of DUDe.

Nevertheless, we still observe very substantial rate gains for DUDe even when compared with biased coupled associations. Detailed breakdowns of these rate gains in various configurations are given in [SinZhaAnd14] and [ElsBoc14], with our findings summarized in Table 2. Here, picocells have transmit power of 30 dBm while femtocells 20 dBm. The gains result mainly from the improved channel quality and also from the biasing as discussed above, which gives cell-edge ($5^{th}$ percentile) and median ($50^{th}$ percentile) UEs access to more resources which results in a UL higher rate. It is quite encouraging that two very different models and approaches to evaluating the rate gains both result in the conclusion that gains in the range of 100-200% are within reach, although the gains do erode somewhat with biasing since the baseline improves. Finally, we note that a recent paper based on optimization theory with a yet different model, also finds significant gains from DUDe [BooBar15].

**Table 2**: Summary of predicted uplink rate gains averaged over all UEs in the network, as a result of DUDe. Picocells have transmit power of 30 dBm while femtocells 20 dBm.

|  | DUDe vs picocells (Bias = 0) | | DUDe vs picocells (Bias = 6 dB) | | DUDe vs femtocells (Bias = 0) | | DUDe vs femtocells (Bias = 8 dB) | |
|---|---|---|---|---|---|---|---|---|
|  | Analysis | Simulation | Analysis | Simulation | Analysis | Simulation | Analysis | Simulation |
| 5$^{th}$ percentile (cell-edge) rate gain | 115% | 90% | 50% | 30% | 270% | 260% | 140% | 95% |
| 50$^{th}$ percentile (Median) rate gain | 95% | 150% | 30% | 60% | 260% | 230% | 120% | 180% |

## 2.4 Different load balancing in the uplink and the downlink

The load that a given BS has in the UL may be different from the load that the same BS may have in the DL. This implies that is is not optimal to have the same set of UEs connected to the same BS in both uplink and the downlink, such that at least some of the UEs should use decoupled access.

Additionally, DUDe allows pushing more UEs to under-utilized small cells in the UL only since it is not limited by interference as is the case in the DL. This results in a better distribution of the UEs among macro and small cells which, in turn, allows for a more efficient resource utilization and higher UL rates (see Figure 3).

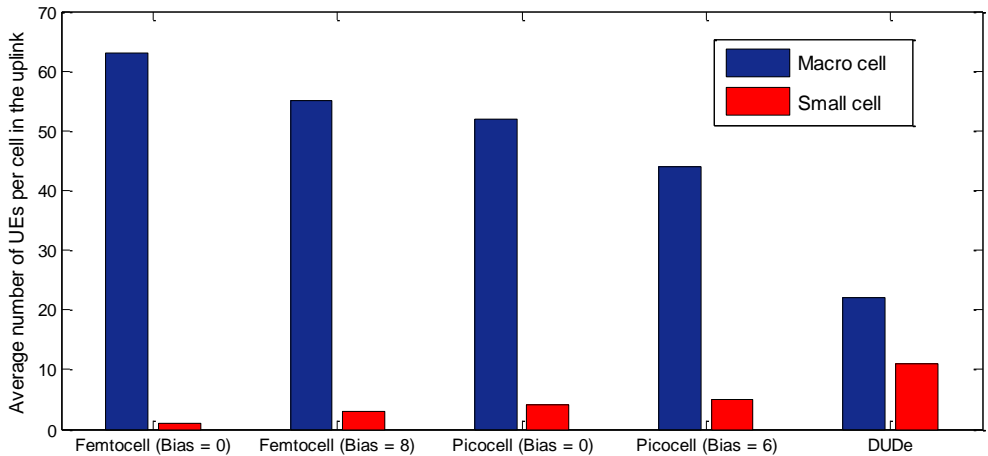

**Figure 3.** Average number of UEs per cell in the UL for the femtocell, picocell and DUDe.

## 2.5 Low deployment costs with RAN centralisation

Implementing a decoupled cell association in a real network requires excellent connectivity and modest cooperation between different base stations. As we will discuss in the subsequent section, the main requirement DUDe imposes is a low latency connection between the downlink and the uplink base stations, to allow a fast exchange of control messages. We emphasise that differently from the most sophisticated forms of CoMP (e.g. joint processing) where a high throughput backhaul connection between BSs is required to allow rapid data exchange, DUDe does not impose a tight requirement on the backhaul capacity. Put another way, DUDe allows gains similar to uplink joint processing (about 100% edge and average throughput gain, as just seen), but with lower deployment costs. Compared to using MIMO or new spectrum to increase the throughput, the cost comparison is even more favourable to DUDe.

The ongoing trend towards using partial or full Radio Access Network (RAN) centralisation in deployments where a high-speed backhaul is available, will be an enabler for downlink and uplink decoupling, as signalling will be routed to a central processing unit with low-latency connections. In particular, partial centralisation refers to those local deployments (e.g. indoor) where the transmission points serving the same local area are all connected to the same baseband processing central unit. Full centralisation, often referred as Cloud-RAN, extends this approach to larger areas, where a large number of RF units are connected to the same baseband processing central unit.

Given this already ongoing trend towards more centralized RAN architectures, which are underpinned by low-latency connectivity between BSs, the incremental cost of DUDe appears negligible in such scenarios.

## 3 DUDe in LTE-A: Enabling Architectures

DUDe can, depending on the deployment scenario and backhaul properties, already be supported by the existing LTE/LTE-A specifications. Illustrated in Figure 4, three specific embodiments are discussed below.

### 3.1 Centralized Processing

As mentioned before, in a deployment scenario with multiple radio units with a different cell-ID connected to a centralized node (e.g. CRAN), DUDe is possible in LTE-A without additional standardization support (see Figure 4(1)). The BS used for downlink transmission to a specific UE is selected using conventional means, typically based on downlink signal strength measurements (possibly with biasing). Uplink transmissions are received by one, or if macro diversity is desirable, multiple radio units as the specifications do not mandate the reception node. Uplink decoding could either be performed at the radio unit (or at the set of radio units) or sampled analog data could be forwarded to the centralized unit via a Common Public Radio Interface (CPRI) interface for further processing.

Uplink-related Access-Stratum (AS) signaling in the downlink, i.e. Layer 1, Layer 2 and RRC control messages related to the uplink transmission sent by the radio unit to the mobile terminal (including e.g. hybrid-ARQ acknowledgements and power-control commands), needs to be transmitted from the downlink node. In the same way, downlink-related AS signaling in the uplink (such as channel quality indicators) needs to be received by the uplink node.

Non-Access Stratum (NAS) signaling, i.e. control messages exchanged between the MME and the mobile terminal (such as establishing and managing bearers, authentication messages, mobility management and tracking area update [OlsRom2009]), can be handled in the same way. Note that NAS signaling has more relaxed requirements on latency.

## 3.2 Shared Cell-ID

An interesting extension of the approach described above is the so-called shared cell-ID approach [ParDah11] (see Figure 4(2)), where radio units all belong to the *same* cell (i.e. have the same cell-ID). Here, CSI enhancements and quasi-co-location mechanisms introduced in Rel-11 as part of the CoMP work are used to rapidly, independently and, from a terminal perspective, transparently switch transmission and reception points for a given terminal. This is a step away from the traditional cell-oriented paradigm towards viewing the antenna points as resources to be used in the best possible way to maximize performance. Furthermore, node association and mobility are handled via proprietary (non-standardized) solutions, transparent to the mobile terminal, and providing better mobility robustness in dense networks compared to methods relying on UE-centric measurements.

Although conceptually straightforward, both centralized processing and shared-ID approaches require a fairly low-latency backhaul to meet the timing requirements of the data plane. Whilst it is possible to have a timing budget in multiples of 8 ms at the cost of an overall performance degradation by exploiting the properties of the uplink HARQ protocol, the de-facto latency requirements are in the range of 3 ms for decoding and scheduling of any retransmissions in LTE. In a practical LTE-A rollout, the deployment is thus limited to remote radio units connected to a centralized baseband processing.

## 3.3 Dual Connectivity

While the two solutions described above require a very low-latency backhaul, usually achieved via connecting radio units (with the same or different cell-ID) to the same central unit, DUDe can also be implemented with a less ideal backhaul. *Dual Connectivity*, an extension first introduced in Rel-12, allows for a terminal to be simultaneously connected to two cells to aggregate data flows or for downlink-uplink decoupling (Figure 4(3)). We note that in Rel-12, DUDe using dual connectivity is limited to inter-frequency deployments, i.e. to deployments where the two cells transmit over different frequency bands; nevertheless, later releases may add support for intra-frequency band deployments. The two cells operate separately, handling their own scheduling and L1/L2 control signaling and thereby significantly relaxing the backhaul requirements compared to the centralized baseband approach. More specifically, the master eNodeB handles downlink data as well as NAS signaling, that is RRC and MME-related signaling, while the secondary eNodeB handles uplink data. This solution has advantages and disadvantages. On one hand a low-latency backhaul connection for the AS signaling is not needed, as AS signaling is terminated in each node. On the other hand, the anchor point for NAS signaling is the MME, which means that node association and mobility must be handled via standardized solutions at the MME side, and proprietary optimization is not possible.

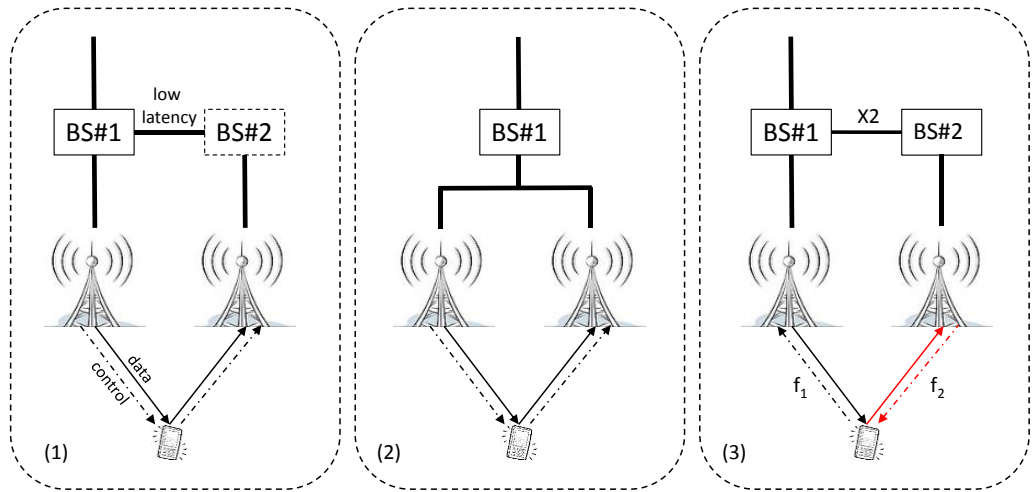

**Figure 4.** The three discussed embodiments of DUDe are: (1) centralized processing unit; (2) shared cell-ID; and (3) the dual connectivity option.

## 4    DUDe in 5G and Beyond

The next few years will see intense research and development on 5G. The ITU is starting their work on requirements, and in 3GPP initial activities on 5G standardization are expected towards the end of 2015 with the overall goal of a large-scale trial around 2018 and commercial operation in 2020. Although any discussion of 5G is by definition speculative, there is an emerging consensus on the data rate requirements and likely key technical features of 5G, including extreme BS densification, massive MIMO, the introduction of millimeter wave bands, and a "cell-less" architecture [And14, BocHea14].

With this view of 5G, in this section we discuss whether 5G (and beyond) standards should include other features to *natively* support DUDe. In other words, we discuss whether a design that is optimized for DUDe from its inception, rather than amended *a posteriori*, could lead to even higher gains.

### 4.1    Major Architectural Changes?

An important question is whether a simple evolution of today's 3GPP architecture design discussed above would be able to efficiently support DUDe in emerging heterogeneous 5G deployments. In the previous section, we discussed how the LTE-A architecture already supports a DUDe implementation when different BSs are connected via fiber to the same radio unit. For the case of different base stations not connected to the same radio unit, we discussed how the support for DUDe in 4G is limited to different frequencies. Any future 5G releases in 3GPP should thus simply allow for a same-frequency dual connectivity, which – despite having implications on resource and interference management – is not considered to be a major upgrade.

A further tweak is needed to ensure proper encryption of all data and control channels, particularly when X2 is used between BSs. However, whilst eNBs can support several (up-to 64 as of Q2 2015) IPsec tunnels per eNB, the management of security via IPsec is so cumbersome that operators tend to deploy only a few IPsec gateways (GWs) per country. Indeed, LTE has seen most IPsec GWs deployed close to the Serving GW (S-GW), which means that traffic logically going via the X2 is actually routed via the S-GW – this incurs a delay which renders DUDe inefficient. Whilst LTE-A enjoys some more IPsec GWs to be deployed closer to the mobile edge, future 5G designs ought to improve security mechanisms and implementations allowing the encryption of X2 traffic without involving the core network.

Given above discussion, however, we conclude that a native support of DUDe does not need major design changes in 5G from an architectural perspective.

### 4.2 DUDe and Hyperdensification

The importance of decoupled selection of the DL/UL access points will grow significantly in the coming years, as 5G will feature hyper-dense deployments in order to meet the high rate demands in crowded spots. One could argue that at extremely high densities of cells, DUDe will lead to lower gains since nearly all the devices will be associated to the nearest small cell in uplink and downlink. However, this will only be true if we assume that all the small cells will have the same power, traffic and deployment characteristics. This is an unrealistic assumption, as future cellular deployments will be characterized by a mixture of user deployed and operator deployed cells, with different power levels, using frequencies ranging from below 1 GHz to tens of GHz, providing services for very different types of traffic and natively supporting device-to-device communications. Therefore, we expect that DUDe gains in future deployments will be even higher w.r.t the ones presented above, especially if we consider the generalized version in which a UE is associated with multiple points and selects the DL or UL direction dynamically, as a part of a scheduling and optimization process.

From a broader perspective, we believe decoupled access necessarily shifts the focus of algorithmic solutions and optimizations towards models that consider two-way traffic from each UE. This is part of a larger trend in wireless network optimization that encompasses full-duplex communication, two-way relaying, and dynamic TDD.

### 4.3 TDD, FDD, or a New Way of Duplexing?

DUDe can work with both FDD and TDD, with different implications from a system-level perspective and from a spectrum-related perspective.

Using TDD allows much more flexibility in trading downlink and uplink resources as compared to FDD. With decoupling, as we have seen, fewer uplink resources are needed to achieve the same uplink rate versus the coupled case, and those resources could be reassigned to the downlink via dynamic TDD, which is in line with the two-way network optimization discussed above. Traditionally, another benefit of TDD is the possibility of estimating the downlink channel via uplink reference signals. This is particularly important for channels with large dimensionality, such as with massive MIMO. Unfortunately, when DUDe is used, downlink and uplink transmissions originate and terminate at different locations, respectively, breaking the channel reciprocity. Much of the existing spectrum is paired FDD spectrum, so for both of these reasons massive MIMO may need to be supported without channel reciprocity, which is an ongoing research topic.

In the medium/long term, DUDe together with different emerging technology trends could require rethinking the traditional FDD/TDD dichotomy. DUDe, hyper densification, the use of higher-frequencies and highly directional antenna arrays, could enable duplexing approaches over the spatial domain. For example, the same band could be used for two different devices, one receiving in downlink from a base station, and the other one transmitting in uplink to another base station. Effectively, assuming an effective downlink/uplink coordinated scheduling mechanism, this could allow full-duplex like performance without the need of using expensive and unproven analog/digital interference-cancellation mechanisms at both base stations and devices.

### 4.4 DUDe with Millimeter Wave Frequencies

Above we discussed why DUDe could make channel estimation via channel reciprocity in TDD more difficult. This effect could be even more pronounced at millimeter wave (mmW) frequencies, where the large number of antenna elements used for beamforming would be enabled by channel reciprocity.

However, there are other factors that point to DUDe as an important enabler for mmW. For example, recent studies on electromagnetic field exposure [ColTho2015] show that to be compliant with applicable exposure limits at frequencies above 6 GHz, the maximum transmit power in the uplink might have to be several dB below the power levels used for current cellular technologies. Since the transmit power has an important impact on uplink coverage (in particular for sounding over a non-precoded channel) we believe a pragmatic approach would be to allocate uplink over a lower frequency with a better link budget. That is, while in the rest of this paper we discussed associating a UE to a macro cell in the downlink and to a small cell in the uplink, for mmW the opposite strategy might prove fruitful: associating the UE to the mmW small cell in the downlink and to a sub 6 GHz macro cell in the uplink.

### 5 Conclusions

In traditional cellular networks, it is practically an axiom that the uplink (UL) connection is always associated with the same Base Station (BS) that has been selected for downlink (DL) reception. In this paper we revise this axiom and introduce the features of Downlink/Uplink

Decoupling (DUDe), a new architectural paradigm where downlink and uplink are not constrained to be associated to the same BS. This is becoming especially relevant in the wake of densification expected in the future cellular networks, where each terminal has multiple access points in proximity. We have identified five key features that demonstrate the usefulness of DUDe, based on a blend of theoretical, experimental, and architectural arguments. We have shown how DUDe can lead to significant gains in network throughput, outage and power consumption at a much lower cost compared to other solutions providing comparable or lower gains. We have discussed the changes needed in the existing LTE-A systems in order to enable DUDe-based operation. We have then presented arguments why DUDe should natively be considered as a part of the future 5G systems. Interestingly, whilst work is needed to have the concept of DUDe in 5G, major changes to the radio access and core networking technologies are not needed. DUDe can be considered as an innovative approach that affects the fundaments of cellular networks and thus opens up a vast opportunity for research and design.

**References**

[ElsBoc14] H. Elshaer, F. Boccardi, M. Dohler and R. Irmer. "Downlink and Uplink Decoupling: a Disruptive Architectural Design for 5G Networks." *IEEE GLOBECOM*, 2014.

[SmiPoGa15] K. Smiljkovikj, P. Popovski, and L. Gavrilovska, "Analysis of the Decoupled Access for Downlink and Uplink in Wireless Heterogeneous Networks", *IEEE Wireless Communications Letters*, early access, January 2015.

[SinZhaAnd14] S. Singh, X. Zhang, and J. G. Andrews, "Joint Rate and SINR Coverage Analysis for Decoupled Uplink-Downlink Biased Cell Associations in HetNets", submitted *IEEE Trans. Wireless Commun.,* Dec. 2014. Available: http://arxiv.org/abs/1412.1898

[BooBha15] H. Boostanimehr and V. Bhargava, "Joint Downlink and Uplink Aware Cell Association in HetNets with QoS Provisioning", submitted to *IEEE Trans. On Wireless Commun.*

[SinAnd14] S. Singh and J. G. Andrews, "Joint Resource Partitioning and Offloading in Heterogeneous Cellular Networks", *IEEE Trans. Wireless Commun.*, vol.13, no.2, pp.888-901, Feb. 2014.

[ParDah11] S. Parkvall, E. Dahlman, G. Jöngren, S. Landström and L. Lindbom Heterogeneous network deployments in LTE. *Ericsson Review*, Feb. 2011.

[And14] J. G. Andrews, S. Buzzi, W. Choi, S. Hanly, A. Lozano, A. CK Soong, and J. C. Zhang "What will 5G be?", *IEEE Journal on Sel. Areas in Commun.*, pp. 1065-82, June 2014.

[Dam11] A. Damnjanovic, J. Montojo, Y. Wei, T. Ji, T. Luo, M. Vajapeyam, T. Yoo, O. Song, and D. Malladi, "A survey on 3GPP heterogeneous networks," *IEEE Wireless Communications,* vol.18, no.3, pp.10-21, June 2011.

[BocHea14] F. Boccardi, R. W. Heath, Jr., A. Lozano, T. L. Marzetta, and P. Popovski "Five disruptive technology directions for 5G," *IEEE Communications Magazine*, Feb. 2014.

[ColTho2015] D. Colombi, B. Thors, and C. Tornevik, "Implications of EMF Exposure Limits on Output Power Levels for 5G Devices above 6 GHz.", *IEEE Antennas & Wireless Propagation Letters,* Feb. 2015.

[OlsRom2009] M. Olsson, S. Rommer, C. Mulligan, S. Sultana and L. Frid. "SAE and the Evolved Packet Core: Driving the mobile broadband revolution," *Academic Press,* 2009.

[AbuAliTaha14] N. Abu-Ali, A.-E.M. Taha, M. Salah and H. Hassanein, "Uplink Scheduling in LTE and LTE-Advanced: Tutorial, Survey and Evaluation Framework," *IEEE Communications Surveys & Tutorials,* vol.16, no.3, pp.1239-1265, 2014.